\documentclass{jfm}

\usepackage{graphicx}
\usepackage{amsmath}
\usepackage{array}
\usepackage{bm}
\usepackage{newtxtext}
\usepackage{newtxmath}
\usepackage{natbib}
\usepackage{hyperref}
\hypersetup{
    colorlinks = true,
    urlcolor   = blue,
    citecolor  = black,
}
\newcommand{\dd}{\mathrm{d}}

\newcolumntype{L}[1]{>{\raggedright\arraybackslash}p{#1}}

\title{A field-equation semi-local transformation for compressible wall turbulence}
\shorttitle{Field-equation semi-local transformation}
\shortauthor{Yin and Wang}

\author{Zifei Yin\aff{1}\corresp{\email{yinzifei@sjtu.edu.cn}}, Jing Wang\aff{1}}

\affiliation{\aff{1}School of Aeronautics and Astronautics, Shanghai Jiao Tong University}

\begin{document}
\maketitle

\begin{abstract}
Compressibility and wall heat transfer change the inner scaling of wall turbulence through the mean density and viscosity fields.
Most semi-local transformations are applied after a wall-normal profile has been selected.
Here the transformed coordinate and transformed velocity are defined by wall-anchored field equations before any profile is extracted.
Wall density and friction velocity enter as boundary data for auxiliary field equations; the resulting reference density and friction velocity, together with the local density and viscosity, set the viscous scaling in the wall layer.
The local logarithmic density contrast is introduced as a bounded correction to the coordinate-stretching factor.
For weak density variation the corrected stretch follows the local inverse kinematic-viscosity variation, while the bounded form limits the influence of finite density contrasts.
The transformed coordinate $Y^+$ and transformed velocity $\bm{U}^+$ are obtained from field equations using the corrected stretching and the mean viscous shear.
In the constant-property limit the density contrast vanishes, $Y^+$ reduces to the ordinary wall coordinate and each transformed velocity component reduces to the conventional mean velocity component in wall units.
For channel flows, the field equations are solved with wall boundary data and the extracted profiles retain their own wall origins.
A cooled shock/boundary-layer interaction examines the response when the wall density and friction velocity vary in the streamwise direction.
Across the cooled high-speed boundary layers considered here, the bounded density correction narrows the inner- and buffer-layer profile spread; symmetric isothermal channels provide the channel comparison with symmetric wall states, whereas mixed thermal-wall channels and the shock interaction expose the residual dependence on unequal wall states and non-equilibrium wall-shear history.
\end{abstract}


\section{Introduction}
\label{sec:introduction}

In high-speed wall turbulence, wall cooling and wall heating produce strong wall-normal variation of the mean density and viscosity across the inner layer \citep{Bradshaw1977,ZhangDuanChoudhari2018}.  The wall units and the mean velocity gradient then depend on Mach number and thermal condition, so incompressible wall units mix property variation with changes in the mean turbulent profile.  The scaling question is how much of this profile variation is removed by ordinary semi-local variables based on local density and viscosity, and what residual inner-layer mismatch remains.

Semi-local transformations address this question by asking whether the compressible mean profile can be measured in a coordinate that accounts for local density and viscosity.  For attached cases with comparable inner-layer state, a successful scaling reduces the spread of the inner and buffer-layer profiles over the Mach numbers and wall temperatures being compared, while any remaining spread identifies physics not carried by the chosen variables.  The transformation is both a probe of the inner momentum scaling and a practical normalisation.  The buffer layer is the most sensitive region, because viscous scaling, property variation and the onset of turbulent transport all act over the same wall-normal range.

Early transformations acted mainly on velocity.  The Van Driest transformation introduced a density-weighted velocity for compressible turbulent boundary layers \citep{VanDriest1951}.  Later semi-local scalings used local density and viscosity to define a wall coordinate that changes with the material properties of the mean flow \citep{HuangColemanBradshaw1995,PatelPeetersBoersmaPecnik2015,PatelBoersmaPecnik2016}.  The Trettel-Larsson (TL) transformation then gave a coordinate-and-velocity transformation for wall turbulence with heat transfer \citep{TrettelLarsson2016}.  The progression is from a velocity correction toward a correction of the wall-normal stretching implied by local material properties.

Mean-stress-balance transformations add another ingredient.  Griffin-Fu-Moin (GFM) used the mean-stress balance to reduce profile spread across compressible boundary layers and channels \citep{GriffinFuMoin2021}; the inverse use of such transformations has also entered near-wall modelling for compressible boundary layers \citep{GriffinFuMoin2023}.  Independent assessments show where the residual mismatch remains: discrepancies often appear in the buffer layer, and the wall thermal condition controls the sign and magnitude of the mismatch after ordinary semi-local scaling based on local density and viscosity \citep{DanisDurbin2024}.

Data-driven transformations have made this residual mismatch more explicit.  \citet{VolpianiIyerPirozzoliLarsson2020} combined near-wall constraints with DNS-fitted power laws for density and viscosity to form a data-driven compressibility transformation.  More recently, \citet{ZhangChenLiaoZhaoXieLiuLu2026} used physics-informed symbolic regression to construct hybrid velocity and temperature transformations, with different effective mappings in different wall-layer regions.  These studies show that the buffer-layer discrepancy contains useful information rather than only scatter between datasets.  The remaining modelling question is what information, beyond local density and viscosity scaling, sets the transformed inner coordinate before the profile is chosen.

The gap addressed here is not another algebraic remapping of a chosen profile, but the definition of the transformed variables before the profile is chosen.  In non-parallel or multi-wall mean flows, the transformed coordinate and velocity are naturally treated as wall-attached field quantities before any station profile is sampled.  The density and viscosity fields set the local viscous length point by point, while the remaining buffer-layer mismatch is represented by a bounded density-induced correction to the coordinate-stretching factor.  Harmonic equations for the density and friction-velocity references, together with Poisson equations for coordinate and velocity, provide the wall anchoring before profiles are extracted, which matters in non-parallel mean flows and in multi-wall flows with unequal thermal boundary conditions.

The field-equation formulation solves harmonic equations for the density and friction-velocity references and Poisson equations for \(Y^+\) and \(\bm{U}^+\) before station profiles are sampled.  The wall fixes the origin, and the wall-distance normal sets the direction in which the coordinate grows.  Differential wall-distance methods show how wall geometry can be encoded by a field equation with explicit boundary conditions \citep{Tucker2003}.  Once the reference fields have been defined, the density contrast supplies a local logarithmic correction to coordinate stretching.  In the uniform-wall limit, ordinary semi-local stretching contains one half-density factor, and the density correction supplies the other half-density factor required to approach a local inverse kinematic-viscosity scale.

The formulation below constructs logarithmic reference fields for density and friction velocity, uses them to set the local viscous scale, and applies a bounded local correction based on the logarithmic density contrast.  The corrected stretching defines \(Y^+\) and, together with the mean viscous shear, defines \(\bm{U}^+\).  On a wall-normal station, the streamwise or wall-tangential component reduces to the profile quantity used for comparison with the Trettel-Larsson and Griffin-Fu-Moin transformations below.  The DNS profiles are thus traces of the field variables or their one-dimensional reductions.

Section \ref{sec:formulation} gives the field-equation formulation and its one-dimensional recovery.  Section \ref{sec:data} summarises the DNS cases and profile comparisons.  The results then examine attached boundary layers, a shock-interaction field and compressible channel flows.  The discussion returns to the range of use and the observed limits of the transformation.

\section{Field-equation formulation}
\label{sec:formulation}

\subsection{Physical idea, notation and required fields}
\label{sec:model_constraints}

In a compressible wall layer, the mean density and viscosity first change the local viscous length.  Ordinary semi-local scaling contains this local material-property effect through a factor proportional to \(\sqrt{\rho}/\mu\).  The remaining buffer-layer mismatch is treated here as a bounded density-induced change in wall-normal coordinate stretching.  The correction is tied to the coordinate-stretching factor and leaves the viscous origin fixed at the no-slip wall.

The notation separates ordinary wall variables from transformed variables.  The mean velocity vector is \(\bm{u}(\bm{x})\).  The ordinary wall coordinate and mean velocity in wall units are denoted by $y^+$ and $\bm{u}^+$; here $\bm{u}^+$ is not a fluctuating velocity.  The transformed coordinate is $Y^+$, and the transformed velocity obtained from the mean viscous shear is denoted by $\bm{U}^+$, with components $U_i^+$.  Thus ordinary profiles are plotted as $(y^+,u^+)$, whereas transformed profiles are plotted as $(Y^+,U^+)$ for the corresponding wall-tangential component.

The formulation uses the mean fields \(\bm{u}\), \(\rho(\bm{x})\) and \(\mu(\bm{x})\) in a selected near-wall region, together with the wall boundary, wall distance, wall-normal direction, any non-wall boundary of the selected region, and the wall density and friction velocity prescribed on the wall.  Reference fields for density and friction velocity set the local viscous scaling and the density contrast.  The density contrast then supplies the logarithmic correction \(\chi\) to coordinate stretching.  The corrected stretching gives \(Y^+\); the same stretching with the mean viscous shear gives \(\bm{U}^+\).

\subsection{Reference fields and local viscous scaling}
\label{sec:density_wall_coordinate}

The field equations are written in physical coordinates on a selected wall-layer mean-flow domain \(\Omega\) with wall boundary \(\Gamma_w\).  In one-dimensional profile reductions \(\Omega\) is the sampled wall-normal interval.  In the two-dimensional boundary-layer and shock-interaction examples it is the wall-layer patch on which the transformed fields are defined and sampled.  In the mixed-channel case \(\Omega\) is the full channel and \(\Gamma_w\) contains both walls.  For a single-wall region, the remaining boundary is \(\Gamma_o=\partial\Omega\setminus\Gamma_w\), with outward unit normal \(\bm{m}\), and is placed outside the near-wall interval over which profiles are compared.  For a domain closed by two wall boundaries, the corresponding non-wall boundary conditions are absent in the wall-normal direction; any additional open boundary or wall-region interface is treated as part of \(\Gamma_o\).

Let $\bm{x}$ be position, and let $\nabla$, $\nabla\cdot$ and \(\nabla^2=\nabla\cdot\nabla\) denote the corresponding gradient, divergence and Laplacian.  The wall boundary fixes the coordinate origin, and the wall-distance normal fixes the direction in which the transformed coordinate grows.  On \(\Gamma_w\), \(\rho_w>0\) denotes the wall density and \(\tau_w>0\) denotes the wall-shear-stress magnitude.  The wall friction velocity is
\begin{equation}
  u_{\tau,w}=\left(\tau_w/\rho_w\right)^{1/2}.
  \label{eq:wall_friction_velocity}
\end{equation}
The wall quantities enter the interior through auxiliary reference fields.  Let \(a_\rho\) and \(a_u\) be logarithmic auxiliary fields for the density and friction-velocity references.  These logarithms are obtained from boundary-value problems so that the reconstructed references remain positive:
\begin{equation}
  \begin{aligned}
  \nabla^2 a_\rho &=0,
  & a_\rho&=\log(\rho_w/\rho_0) &&\hbox{on }\Gamma_w,
  & \nabla a_\rho\boldsymbol{\cdot}\bm{m}&=0 &&\hbox{on }\Gamma_o,\\
  \nabla^2 a_u &=0,
  & a_u&=\log(u_{\tau,w}/u_{\tau,0}) &&\hbox{on }\Gamma_w,
  & \nabla a_u\boldsymbol{\cdot}\bm{m}&=0 &&\hbox{on }\Gamma_o,
  \end{aligned}
  \label{eq:reference_field_pdes}
\end{equation}
The constants \(\rho_0>0\) and \(u_{\tau,0}>0\) are arbitrary positive reference constants introduced to make the logarithms dimensionless.  The physical reference fields are
\begin{equation}
  \rho_r=\rho_0\exp(a_\rho),\qquad
  u_{\tau,r}=u_{\tau,0}\exp(a_u),\qquad
  \tau_r=\rho_r u_{\tau,r}^2 .
  \label{eq:reference_fields}
\end{equation}
The constants \(\rho_0\) and \(u_{\tau,0}\) set gauges for \(a_\rho\) and \(a_u\): changing them shifts the logarithmic fields by constants but leaves \(\rho_r\), \(u_{\tau,r}\), \(\tau_r\) and the transformation unchanged.  The physical wall information is the Dirichlet data in (\ref{eq:reference_field_pdes}); after these boundary-value problems are solved, the interior scaling uses the local values of \(\rho_r\) and \(u_{\tau,r}\).

The inverse local viscous length used in the field equations is
\begin{equation}
  G=\frac{\sqrt{\rho\,\rho_r}\,u_{\tau,r}}{\mu},
  \label{eq:g_def}
\end{equation}
where $\rho$ and $\mu$ are the mean density and dynamic viscosity.  Equivalently, \(G=\sqrt{\rho\,\tau_r}/\mu\) because \(\tau_r=\rho_r u_{\tau,r}^2\).  Thus \(G\) combines local material properties with the solved density and friction-velocity references.  For the wall-specific coordinate being constructed, let $d_w(\bm{x})$ be the distance to that wall and
\begin{equation}
  \bm{n}=\frac{\nabla d_w}{|\nabla d_w|}
  \label{eq:normal}
\end{equation}
the unit wall-distance normal, directed away from that wall.  Thus $\bm{n}$ is the local direction of wall-normal coordinate growth for the chosen wall, whereas $\bm{m}$ is the outward normal on any non-wall boundary used to close the selected region.

The density contrast used in the coordinate-stretch correction is measured relative to the density reference field,
\begin{equation}
  S_\rho=\log(\rho/\rho_r).
  \label{eq:srho_gauge}
\end{equation}
The factor $\frac{1}{2}S_\rho$ is the local logarithmic density term after the reference field has been defined.  The viscosity part is already contained in the baseline factor \(G\) and is not repeated in the density term below.  For a uniform-wall case this gives \(S_\rho=\log(\rho/\rho_w)\); for a channel with unequal wall data, \(\rho_r\) is the full-channel reference field obtained from the two wall boundary values.

\subsection{Density-induced coordinate correction}
\label{sec:coordinate_correction}

The logarithmic density term entering the coordinate-stretch correction is
\begin{equation}
  q=\frac{S_\rho}{2}.
  \label{eq:q_single_source}
\end{equation}
The bounded density correction is
\begin{equation}
  \chi=C(q)=q_0\tanh(q/q_0),
  \qquad q_0=0.20 .
  \label{eq:bounded_source}
\end{equation}
Thus \(q\) is the local half-density logarithm relative to the solved reference density, and \(\chi\) is the finite-amplitude correction applied to the coordinate-stretching factor.  The bounded response leaves weak density changes essentially linear, since \(C(q)=q+O(q^3)\), but prevents very large density contrasts from dominating the stretching.  It does not alter the density field or introduce wall values into interior points.

The corrected inverse viscous length used by the coordinate and velocity equations is
\begin{equation}
  G_Y=G\exp\chi .
  \label{eq:corrected_stretching}
\end{equation}
For a single-wall or uniform-wall case, the unbounded weak-density form has a direct interpretation:
\begin{equation}
  G\exp q
  =
  \frac{\sqrt{\rho\,\rho_r}\,u_{\tau,r}}{\mu}
  \left(\frac{\rho}{\rho_r}\right)^{1/2}
  =
  \frac{\rho u_{\tau,r}}{\mu}
  =
  \frac{u_{\tau,r}}{\nu}.
  \label{eq:kinematic_viscosity_limit}
\end{equation}
The density correction therefore supplies the half-density exponent that is absent from the baseline semi-local factor \(G\), moving the coordinate stretching toward a local inverse kinematic-viscosity scale.  Equation (\ref{eq:bounded_source}) keeps this mechanism linear for weak density variation and bounded for finite density contrasts.

The weak-property-gradient limit makes the role of the density term explicit.  For uniform wall reference data, let
\[
  \rho=\rho_w(1+\epsilon \rho_1),\qquad
  \mu=\mu_w(1+\epsilon \mu_1),\qquad
  \epsilon\ll1 .
\]
Then
\[
  G=
  \frac{\sqrt{\rho_w\tau_w}}{\mu_w}
  \left[
  1+\epsilon\left({\textstyle\frac{1}{2}}\rho_1-\mu_1\right)
  \right]+O(\epsilon^2),
  \qquad
  \chi=q+O(\epsilon^3)=\epsilon{\textstyle\frac{1}{2}}\rho_1+O(\epsilon^2),
\]
and therefore
\begin{equation}
  G_Y=
  \frac{\sqrt{\rho_w\tau_w}}{\mu_w}
  \left[
  1+\epsilon\left(\rho_1-\mu_1\right)
  \right]+O(\epsilon^2).
  \label{eq:weak_density_limit}
\end{equation}
To leading order the corrected stretching follows the local inverse kinematic-viscosity variation.  The wall origin is still fixed by the boundary condition for \(Y^+\), not by imposing a separate condition on \(\chi\).

\subsection{Coordinate and velocity fields}
\label{sec:final_projection}

The transformed coordinate is obtained as a field whose normal derivative is tied to the corrected stretching:
\begin{equation}
  \nabla^2Y^+
  =
  \nabla\cdot
  \left(G_Y\bm{n}\right),
  \qquad
  Y^+=0
  \quad \hbox{on the wall}.
  \label{eq:ytrans_pde}
\end{equation}
Since $\nabla$ is taken with respect to dimensional physical coordinates, \(G_Y\) is an inverse length.  The wall condition applies to the wall whose coordinate is being defined; in a multi-wall domain, each wall-specific coordinate has its own wall origin and wall-distance normal.  The Poisson form lets neighbouring wall-normal lines share a compatible multidimensional coordinate before any station profile is extracted.
At any non-wall outer boundary or interface with a neighbouring wall region, the normal component of \(\nabla Y^+\) is matched to the corrected coordinate gradient:
\begin{equation}
  \left[
    \nabla Y^+-G_Y\bm{n}
  \right]\boldsymbol{\cdot}\bm{m}=0 .
  \label{eq:ytrans_bc}
\end{equation}
In a single-wall region the reference fields, \(\chi\), \(Y^+\) and \(\bm{U}^+\) are defined on the same wall layer.  In a multi-wall region, the reference fields may be solved on the larger domain, while the transformed coordinate and velocity remain wall-specific: for each wall \(k\), \(Y_k^+=0\) on that wall and the growth direction is set by its wall-distance normal.

In component equations, \(x_j\) is the \(j\)th physical coordinate, \(\partial_j=\partial/\partial x_j\), and repeated Roman indices imply summation.  The mean viscous-shear vector used to define the transformed velocity is
\begin{equation}
  \tau^\dagger_{v,i}
  =
  \frac{\mu n_j\partial_j u_i}{\tau_r},
  \label{eq:vector_tauv}
\end{equation}
where \(u_i\) is the \(i\)th mean velocity component.  The vector $\bm{\tau}_v^\dagger$ is the wall-normal viscous derivative of the mean velocity vector scaled by the shear reference \(\tau_r=\rho_r u_{\tau,r}^2\).  The transformed velocity field is then defined component by component:
\begin{equation}
  \nabla^2 U_i^+
  =
    \frac{\partial}{\partial x_j}
  \left[
    G_Y n_j
    \tau^\dagger_{v,i}
  \right],
  \qquad
  U_i^+=0
  \quad \hbox{on the wall}.
  \label{eq:utrans_pde}
\end{equation}
The corresponding condition on any non-wall outer boundary or interface is
\begin{equation}
  \left[
    \frac{\partial U_i^+}{\partial x_j}
    -
    G_Y n_j
    \tau^\dagger_{v,i}
  \right]m_j=0 .
  \label{eq:projection_bc}
\end{equation}
The component form in (\ref{eq:utrans_pde}) states the divergence convention used for the transformed velocity field.  In a multi-wall region the same corrected stretching can therefore be used to form several velocity profiles, each defined with respect to its own wall origin and shear scale.  In the canonical profile comparisons, the scalar shear reduces to the streamwise or wall-tangential viscous shear normalised by the wall shear in the uniform-wall limit.  The one-dimensional profile reduction of these field equations is given in \S\ref{sec:one_dimensional_reduction}.

For comparison with wall-turbulence inner laws, the wall-parallel velocity component is obtained from
\begin{equation}
  \bm{u}_t=\bm{P}\bm{u},\qquad
  \bm{P}=\bm{I}-\bm{n}\bm{n} ,
  \label{eq:tangent_projector}
\end{equation}
where $\bm{u}_t$ is the wall-tangential part of the mean velocity, $\bm{P}$ is the tangent-plane projector and $\bm{I}$ is the identity tensor.  The dyadic product $\bm{n}\bm{n}$ removes the wall-normal velocity component.  The wall origin and coordinate direction are set by the boundary conditions for \(Y^+\) and the wall-distance normal; the interior scaling is evaluated from the reference fields and the local density correction.

\subsection{One-dimensional recovery}
\label{sec:one_dimensional_reduction}

For a one-dimensional profile with a fixed wall footpoint, let \(y\) be the dimensional wall-normal coordinate.  Equation (\ref{eq:ytrans_pde}) reduces to
\begin{equation}
  \frac{\dd Y^+}{\dd y}=G_Y=G\exp\chi .
  \label{eq:coupled_1d}
\end{equation}
The transformed coordinate and velocity are anchored by \(Y^+=0\) and \(U^+=0\) at the wall.  In deriving (\ref{eq:coupled_1d}), the integration constant is zero because the one-dimensional form of the outer or interface condition (\ref{eq:ytrans_bc}) imposes the same normal derivative there.  Thus the field equation recovers ordinary profile integration when a single wall-normal line has already been selected.  For uniform wall data on the selected wall region, the reference-field equations have the exact constant solution \(\rho_r=\rho_w\) and \(u_{\tau,r}=u_{\tau,w}\).  In that special case
\begin{equation}
  \rho_r=\rho_w,\qquad
  u_{\tau,r}=u_{\tau,w},\qquad
  \tau_r=\tau_w,\qquad
  G=\frac{s}{\delta_{\nu,w}},\qquad
  S_\rho=\log(\rho/\rho_w),
  \label{eq:uniform_wall_reduction}
\end{equation}
where \(\rho^+=\rho/\rho_w\), \(\mu^+=\mu/\mu_w\), \(s=\sqrt{\rho^+}/\mu^+\), \(u_{\tau,w}=(\tau_w/\rho_w)^{1/2}\), \(\nu_w=\mu_w/\rho_w\) and \(\delta_{\nu,w}=\nu_w/u_{\tau,w}=\mu_w/\sqrt{\rho_w\tau_w}\).  Equivalently,
\begin{equation}
  \log G=\log(1/\delta_{\nu,w})+\frac{1}{2}\log(\rho/\rho_w)-\log(\mu/\mu_w).
  \label{eq:log_g_uniform}
\end{equation}
Equations (\ref{eq:uniform_wall_reduction}) and (\ref{eq:log_g_uniform}) apply when the reference-field equations have constant wall solutions.  With \(y_w\) as the wall location and \(y^+=(y-y_w)/\delta_{\nu,w}\), the coordinate equation becomes
\begin{equation}
  \frac{\dd Y^+}{\dd y^+}=s\exp\chi .
  \label{eq:profile_coordinate_limit}
\end{equation}
The wall-tangential velocity component used in the profile figures follows from the same corrected stretching.  In one dimension, (\ref{eq:utrans_pde}) gives
\begin{equation}
  \frac{\dd U^+}{\dd y}=\tau_v^\dagger G\exp\chi,
  \qquad
  \frac{\dd U^+}{\dd Y^+}=\tau_v^\dagger ,
  \label{eq:profile_limit}
\end{equation}
where \(\tau_v^\dagger\) is the wall-normal viscous shear scaled by \(\tau_r\).  In the uniform-wall profile notation, \(\tau_v^\dagger=\tau_v^+\), with \(\tau_v^+=\mu\,\dd u_t/\dd y\,/\,\tau_w\) for the chosen wall-tangential component, and therefore
\begin{equation}
  \frac{\dd U^+}{\dd Y^+}=\tau_v^+ .
  \label{eq:transformed_shear}
\end{equation}
The velocity equation integrates the measured mean viscous shear along the transformed wall coordinate supplied by the coordinate equation.  These are the one-dimensional traces of the field equations.

The constant-property limit is the identity transformation.  If $\rho=\rho_w$, $\mu=\mu_w$ and \(\tau_w\) is constant, the reference-field equations give constant \(\rho_r=\rho_w\), \(u_{\tau,r}=u_{\tau,w}\) and \(\tau_r=\tau_w\).  Then \(G=1/\delta_{\nu,w}\), \(S_\rho=0\), \(q=0\), \(\chi=0\) and $Y^+=y^+$.  Equation (\ref{eq:profile_limit}) integrates the viscous shear in wall units.  For a velocity component with zero wall value, the transformed profile is identical to the velocity profile in wall units.

This one-dimensional reduction also clarifies the distinction from standard profile transformations.  TL uses the local density and viscosity variation along the sampled wall-normal line to define the semi-local coordinate and velocity correction.  GFM instead uses the mean-stress balance to set the transformed inner shear.  In the present reduction the shear relation is \(\dd U^+/\dd Y^+=\tau_v^+\), while the coordinate-stretching factor is \(s\exp\chi\); the additional factor \(\chi\) is the bounded density correction associated with the logarithmic density contrast.

\begin{figure}
  \centerline{\includegraphics[width=0.98\textwidth]{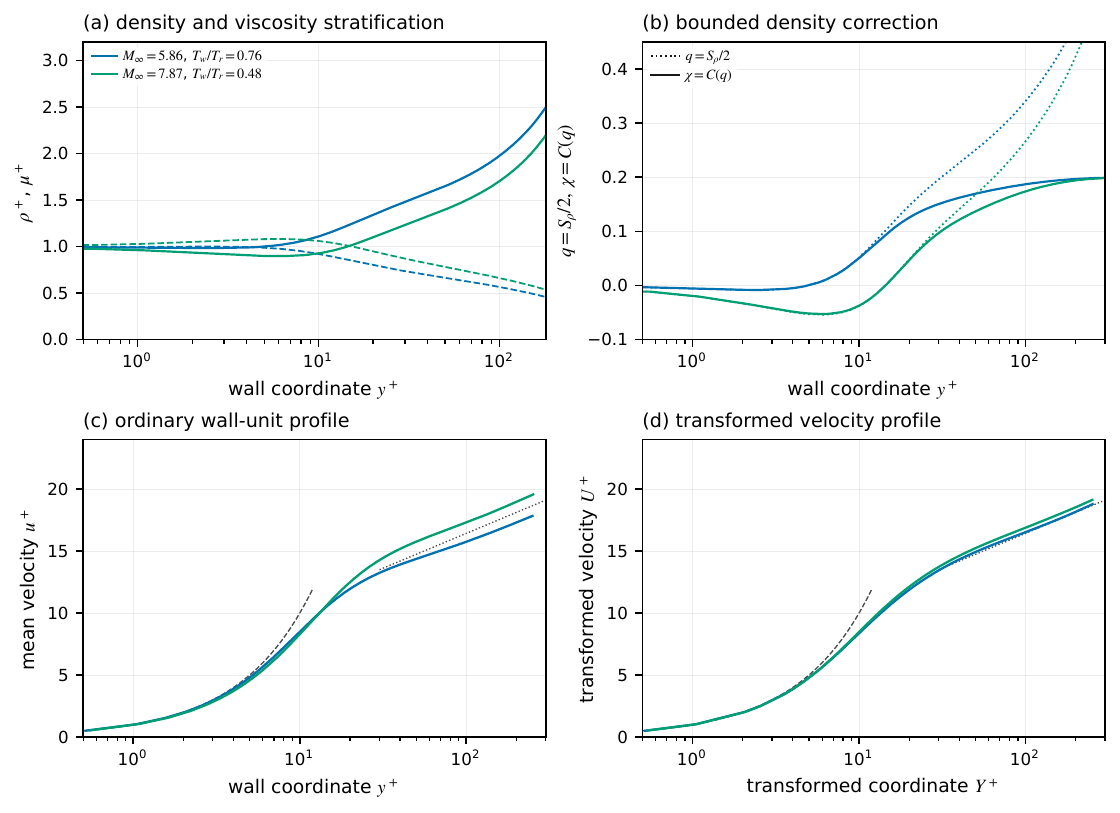}}
  \caption{Density-stratified coordinate transformation for two cooled Zhang-Duan-Choudhari flat plates \citep{ZhangDuanChoudhari2018}.  These are uniform-wall cases, so the reference-field equations have the constant solution \(\rho_r=\rho_w\), \(u_{\tau,r}=u_{\tau,w}\), \(\tau_r=\tau_w\) and \(G=s/\delta_{\nu,w}\).  (a) Wall-normal profiles of $\rho^+=\rho/\rho_w$ and $\mu^+=\mu/\mu_w$ that form the ordinary semi-local stretch $s=\sqrt{\rho^+}/\mu^+$; solid curves denote \(\rho^+\) and dashed curves denote \(\mu^+\).  (b) Logarithmic density term \(q=S_\rho/2\) (dotted) and bounded correction \(\chi=C(q)\) (solid).  (c) Mean velocity profiles in ordinary wall units $(y^+,u^+)$.  (d) Transformed velocity profiles obtained from the mean viscous shear, plotted as $(Y^+,U^+)$.  Dashed and dotted black curves denote the viscous law and log law.}
  \label{fig:metric_mechanism}
\end{figure}

Figure \ref{fig:metric_mechanism} follows this one-dimensional reduction for two cooled Zhang-Duan-Choudhari profiles \citep{ZhangDuanChoudhari2018}.  These uniform-wall profiles use the constant reference-field solution, so the density contrast reduces to the usual wall-normalised form and \(G=s/\delta_{\nu,w}\).  The panels show how the ordinary semi-local stretch, the logarithmic input \(q=S_\rho/2\), the bounded correction \(\chi=C(q)\), and the mean viscous shear combine to give the transformed profiles.

\subsection{Mean fields and wall boundary data}
\label{sec:transformation_steps}

The transformation is specified by the mean velocity, density and viscosity in the selected near-wall region; the wall boundary, wall distance and wall-normal direction; the non-wall boundary or wall-region interface; and the wall density and friction velocity used as boundary data for the reference-field equations.  Together these quantities define the reference fields \(\rho_r\) and \(u_{\tau,r}\), the derived shear scale \(\tau_r\), the inverse viscous length \(G\), the density contrast \(S_\rho\), the logarithmic density term \(q\), the bounded correction \(\chi=C(q)\), the transformed coordinate \(Y^+\) and the transformed velocity \(\bm{U}^+\).

For the attached boundary-layer profiles considered here, the wall data are uniform on the selected wall region and the reference-field equations reduce to constant \(\rho_r=\rho_w\), \(u_{\tau,r}=u_{\tau,w}\) and \(\tau_r=\tau_w\).  When published DNS statistics are available only in station-based wall units, the corresponding uniform-wall representation is
\begin{equation}
  S_\rho(x,y)=\log\rho^+(x,y),\qquad
  G(x,y)=\alpha(x)\frac{\sqrt{\rho^+(x,y)}}{\mu^+(x,y)},\qquad
  \alpha(x)=\frac{\partial y^+}{\partial y}=\frac{1}{\delta_{\nu,w}(x)} ,
  \label{eq:local_wall_unit_field_form}
\end{equation}
where \(\rho^+\) and \(\mu^+\) are normalised by the uniform wall values for that station and \(\alpha\) is the inverse wall viscous length in the published statistics.  This representation is the uniform-wall reduction of (\ref{eq:g_def}) and (\ref{eq:srho_gauge}).  For cases with wall data varying along the surface, such as the shock-interaction field below, the reference fields are obtained from (\ref{eq:reference_field_pdes}) using the wall-surface data as boundary conditions.  The transformed velocity is obtained from the mean viscous shear supplied by the mean velocity and viscosity fields, and the plotted profiles are sampled from the resulting transformed coordinate and velocity fields.

\section{DNS cases and profile comparisons}
\label{sec:data}

All profile comparisons use the logarithmic density term introduced in \S\ref{sec:coordinate_correction}.  The constant \(q_0=0.20\) is kept fixed for all cases, with \(q=S_\rho/2\) and \(\chi=C(q)=q_0\tanh(q/q_0)\).  The DNS cases, wall boundary data and boundary-layer spread measure are summarised in this section.

The two cooled zero-pressure-gradient (ZPG) flat plates from \citet{ZhangDuanChoudhari2018} shown in figure \ref{fig:metric_mechanism}, at $M_\infty=5.86$, $T_w/T_r=0.76$ and $M_\infty=7.87$, $T_w/T_r=0.48$, illustrate the one-dimensional uniform-wall reduction.  Here \(M_\infty\) is the free-stream Mach number, \(T_w\) is the wall temperature and \(T_r\) is the recovery temperature.  These external boundary layers are uniform-wall special cases, so the reference-field equations have constant solutions and \(S_\rho=\log(\rho/\rho_w)\).  Their strong cold-wall density and viscosity stratification exposes the action of \(\chi=C(q)\) in figure \ref{fig:metric_mechanism}(b,d).

The remaining cases cover attached boundary layers, internal flows and a non-equilibrium interaction.  The rest of the Zhang-Duan-Choudhari profiles sample the same high-Mach cold-wall family, the Volpiani-Bernardini-Larsson Mach 2.28 ZPG data \citep{VolpianiBernardiniLarsson2018,Zenodo8307822} give station-resolved boundary-layer development at fixed Mach number, and the Cogo profiles \citep{CogoBauChinappiBernardiniPicano2023,CogoSalvadorePicanoBernardini2023} extend the attached-boundary-layer evidence to a wider heat-transfer range at higher Reynolds number.  The Huang-Coleman-Bradshaw and Gerolymos-Vallet data \citep{HuangColemanBradshaw1995,GerolymosVallet2023,GerolymosVallet2024} examine symmetric isothermal channel flows.  The Lusher-Coleman mixed thermal-wall channels impose unequal thermal wall states, so the density and friction-velocity reference fields are solved on the full channel with two different wall boundary conditions.  The Hirai-Kawai shock-interaction field \citep{HiraiKawai2023} introduces streamwise-varying wall data in a cooled-wall SBLI.

The DNS flow conditions are summarised in table \ref{tab:dns_flow_conditions}.  The selected mean fields and wall boundary data supply the wall-shear and material-property information for the transformation.  The boundary-layer cases span $M_\infty=2.0$-$13.64$, include the cooled and nearly adiabatic flat plates of \citet{ZhangDuanChoudhari2018}, and add the Cogo high-speed boundary layers \citep{CogoBauChinappiBernardiniPicano2023,CogoSalvadorePicanoBernardini2023}.  In the table, \(T_\infty\) is the free-stream temperature, \(Re_\theta\) is the momentum-thickness Reynolds number and \(Re_\tau\) is the friction Reynolds number.  The Volpiani-Bernardini-Larsson Mach 2.28 data contain 121 developing ZPG profiles, with \(x/\delta_i\) denoting streamwise distance normalised by the inlet boundary-layer thickness \(\delta_i\).  For the shock-interaction case, \(x_{sh}\) is the nominal shock-impingement location and \(L_{int}\) is the interaction length defined by \citet{HiraiKawai2023}.

\begin{table}
  \begin{center}
  \small
  \setlength{\tabcolsep}{3.2pt}
  \begin{tabular}{L{0.22\textwidth}L{0.18\textwidth}L{0.16\textwidth}L{0.24\textwidth}L{0.12\textwidth}}
  \hline
  DNS study & Flow configuration & Thermal state & Range & Wall-layer feature \\
  \hline
  \citet{ZhangDuanChoudhari2018} & Attached high-speed ZPG boundary layers & Cooled to nearly adiabatic walls, $T_w/T_r=0.18$-$1.00$ & $M_\infty=2.5$-$13.64$; $Re_\theta=2.1\times10^3$-$1.44\times10^4$ & Cold-wall density stratification \\
  \citet{CogoBauChinappiBernardiniPicano2023,CogoSalvadorePicanoBernardini2023} & Isothermal-wall ZPG boundary layers & $T_w/T_\infty=1.18$-$7.45$ & $M_\infty=2.0$-$6.0$; $Re_\tau\simeq443$-$1947$ & Extended heat-transfer range \\
  Volpiani-Bernardini-Larsson Mach 2.28 station-resolved data \citep{VolpianiBernardiniLarsson2018,Zenodo8307822} & Developing ZPG boundary layers & $T_w/T_r=1.00,1.90$ & $x/\delta_i\simeq20$-$60$; 121 stations & Streamwise field variation \\
  \citet{HuangColemanBradshaw1995,GerolymosVallet2023,GerolymosVallet2024} & Symmetric isothermal channels & Isothermal walls & Centreline Mach number $0.32$-$3.0$; $Re_\tau\simeq107$-$1479$ & Symmetric isothermal walls \\
  \citet{LusherColeman2023} & Mixed thermal-wall channels & Isothermal and adiabatic walls & Friction-scaled Mach number $0.0955$-$0.20$; friction Reynolds number $193.9$-$3937.5$ & Mixed thermal walls \\
  \citet{HiraiKawai2023} & Cooled-wall shock/boundary-layer interaction & $T_w/T_r=0.5$ & $M_\infty=2.28$; $(x-x_{sh})/L_{int}=-3.44$-$4.31$ & Wall-shear recovery \\
  \hline
  \end{tabular}
  \caption{DNS flow conditions for the transformation comparisons.  Reynolds and Mach numbers follow the notation of the original DNS studies; \(\delta_i\), \(x_{sh}\) and \(L_{int}\) are defined in the text.  The final column lists the wall-layer variation or channel configuration represented by each DNS set.}
  \label{tab:dns_flow_conditions}
  \end{center}
\end{table}

The profile comparisons use Trettel-Larsson and GFM as baselines, evaluated on the same DNS profiles as the present transformation.  The viscous line and the classical logarithmic law, $U^+=\kappa^{-1}\log Y^+ + 5.2$, provide reference curves.  For the boundary-layer profile sets, the spread is reported as \(\Delta_{\mathrm{buf}}\), the mean envelope width of the plotted profiles over \(5\le Y^+\le70\) after interpolation to a common \(Y^+\) grid.  The values for figures \ref{fig:tbl_validation} and \ref{fig:independent_high_speed_bl} are collected in table \ref{tab:buffer_spread}.

\section{Boundary-layer transformations and shock-interaction recovery}
\label{sec:results}

\subsection{Boundary-layer collapse and transformed coordinates}
\label{sec:tbl_comparison}

\begin{figure}
  \centerline{\includegraphics[width=0.98\textwidth]{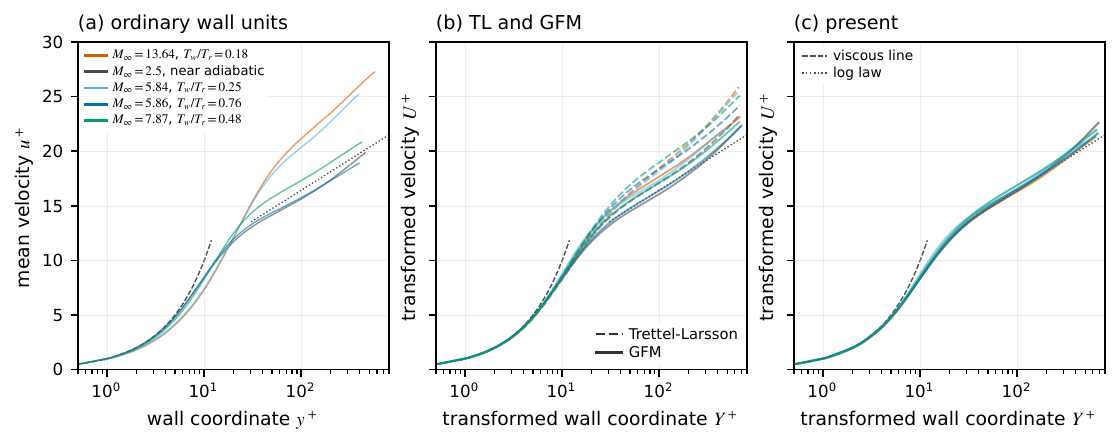}}
  \caption{Mean-profile collapse for the Zhang-Duan-Choudhari smooth high-speed ZPG flat plates \citep{ZhangDuanChoudhari2018}.  (a) Mean velocity profiles in ordinary wall units $(y^+,u^+)$.  (b) The same profiles after Trettel-Larsson and GFM transformations, plotted as transformed velocity $U^+$ versus transformed coordinate $Y^+$; dashed and solid coloured curves denote Trettel-Larsson and GFM, respectively.  (c) Profiles after the present transformation, plotted as $(Y^+,U^+)$.  Colours in all panels identify Mach number and wall thermal condition.  Dashed and dotted black curves denote the viscous law and log law.}
  \label{fig:collapse_overview}
\end{figure}

Figure \ref{fig:collapse_overview} compares the profile collapse for the Zhang-Duan-Choudhari flat plates.  In ordinary wall units, changes in Mach number and wall temperature produce a broad family of mean profiles.  Trettel-Larsson and GFM reduce part of this spread.  For the cases shown, the present transformation gives a narrower inner- and buffer-layer band, while the outer part of the profiles retains case dependence.

\begin{figure}
  \centerline{\includegraphics[width=0.98\textwidth]{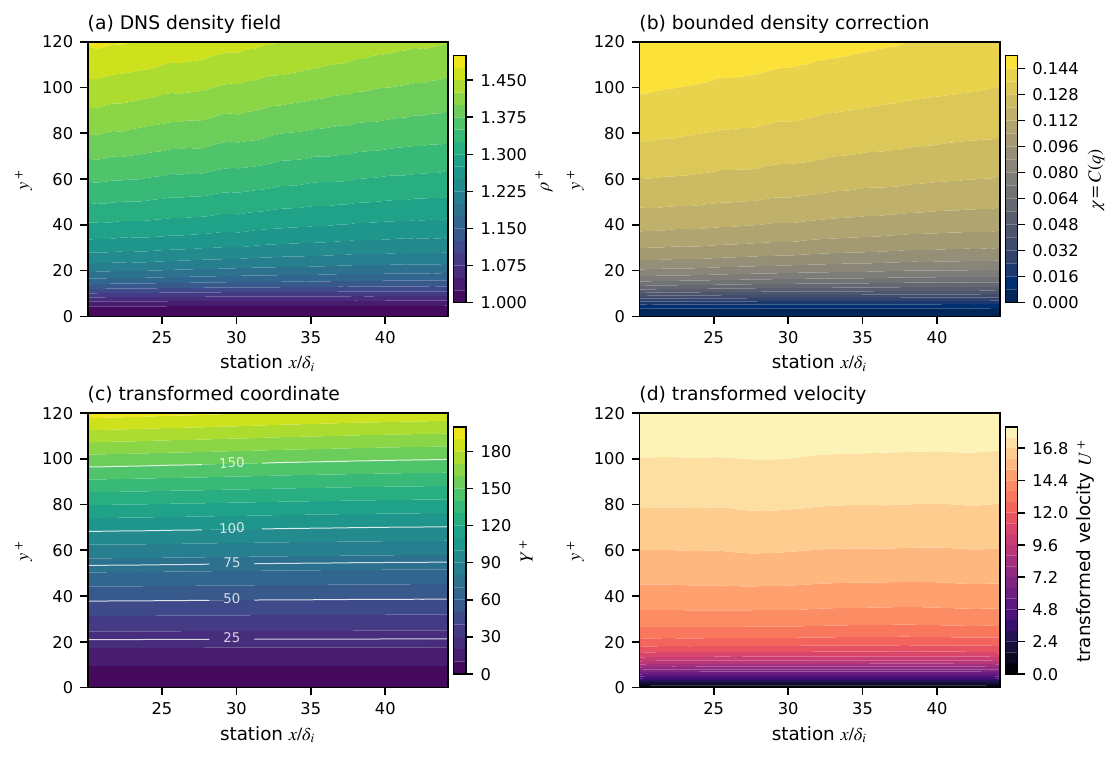}}
  \caption{Two-dimensional boundary-layer fields for the Mach 2.28 ZPG mean field from the Volpiani-Bernardini-Larsson data \citep{VolpianiBernardiniLarsson2018,Zenodo8307822}, plotted against streamwise station $x/\delta_i$ and wall-normal distance $y^+$.  Here \(\delta_i\) is the inlet boundary-layer thickness of the DNS, \(y^+\) is the ordinary wall coordinate, and \(\rho^+=\rho/\rho_w\).  (a) Mean density field $\rho^+$.  (b) Bounded density correction \(\chi=C(q)\).  (c) Transformed coordinate $Y^+$ with selected coordinate contours.  (d) Transformed streamwise velocity $U^+$.  The lower boundary is the wall; the wall condition, wall-distance normal and outer-boundary flux conditions define the transformed coordinate and velocity from which station profiles are sampled.}
  \label{fig:field_patch}
\end{figure}

Figure \ref{fig:field_patch} shows the uniform-wall reduction on a two-dimensional Mach 2.28 ZPG mean field from the Volpiani-Bernardini-Larsson data.  The density varies weakly in the streamwise direction but strongly across the inner layer.  The local bounded correction \(\chi=C(q)\) changes the coordinate-stretching factor; the transformed coordinate contours and transformed velocity are then obtained on the wall-layer field before station profiles are sampled.

\subsection{Boundary-layer transformation comparison}
\label{sec:tbl_method_comparison}

\begin{figure}
  \centerline{\includegraphics[width=0.98\textwidth]{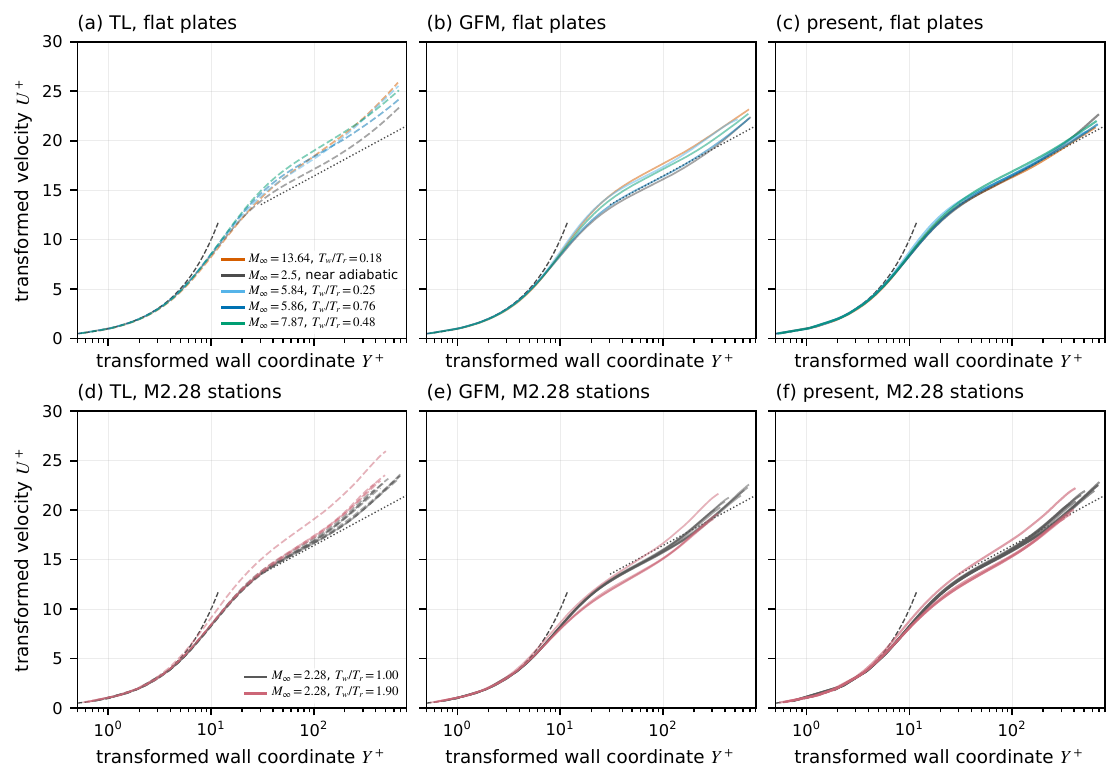}}
  \caption{Boundary-layer transformation comparison.  The columns correspond to Trettel-Larsson, GFM and the present transformation.  (a-c) Attached high-speed ZPG flat-plate profiles from \citet{ZhangDuanChoudhari2018}.  (d-f) Station-resolved Mach 2.28 ZPG boundary-layer profiles from the Volpiani-Bernardini-Larsson data \citep{VolpianiBernardiniLarsson2018,Zenodo8307822}.  Profiles are plotted as transformed velocity versus transformed coordinate.  Colours identify the Mach number and wall thermal condition of the Zhang-Duan-Choudhari cases in the upper row; in the lower row grey and pink curves denote the \(T_w/T_r=1.00\) and \(T_w/T_r=1.90\) station groups, respectively.  Dashed and dotted black curves denote the viscous law and log law.}
  \label{fig:tbl_validation}
\end{figure}

Figure \ref{fig:tbl_validation} compares the three transformations for the Zhang-Duan-Choudhari flat plates and the station-resolved Mach 2.28 ZPG boundary layers from the Volpiani-Bernardini-Larsson data.  In the Zhang-Duan-Choudhari row, the present transformation gives the smallest buffer-layer envelope among the three transformations in table \ref{tab:buffer_spread}.  Trettel-Larsson and GFM still reduce the wall-unit spread, but leave larger cold-wall buffer-layer offsets in this set.

In the lower row of figure \ref{fig:tbl_validation}, the Mach 2.28 station sequence is a developing boundary layer with weaker property stratification.  GFM and the present transformation give comparable buffer-layer envelopes in this set.  The difference between them is small because the density-induced correction to the coordinate-stretching factor is a modest addition to the ordinary semi-local stretching.  The three columns separate the mean-field quantities used by the transformations: TL follows the local density and viscosity, GFM follows the mean-stress balance and the present transformation changes the coordinate stretching through the bounded density contrast.

\subsection{Independent high-speed boundary layers}
\label{sec:independent_validation}

The Cogo DNS profiles of \citet{CogoBauChinappiBernardiniPicano2023,CogoSalvadorePicanoBernardini2023} provide a second attached-boundary-layer data set.  These profiles cover a wider heat-transfer range at $M_\infty=2$-$6$ and come from a DNS family independent of the Zhang-Duan-Choudhari flat plates used above.

\begin{figure}
  \centerline{\includegraphics[width=0.98\textwidth]{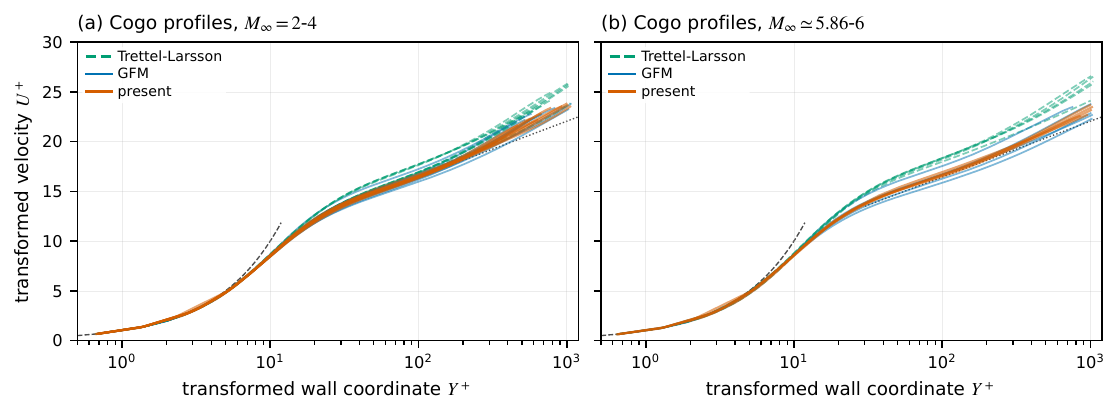}}
  \caption{Cogo high-speed ZPG boundary-layer comparison \citep{CogoBauChinappiBernardiniPicano2023,CogoSalvadorePicanoBernardini2023}.  (a) $M_\infty=2$-$4$ cases.  (b) $M_\infty\simeq5.86$-$6$ cases.  Each panel compares Trettel-Larsson (green dashed), GFM (blue solid) and the present transformation (orange solid), with thin curves denoting individual Cogo profiles within each Mach-number group.  Profiles are plotted as transformed velocity versus transformed coordinate.  Dashed and dotted black curves denote the viscous law and log law.}
  \label{fig:independent_high_speed_bl}
\end{figure}

Across the Cogo profiles in figure \ref{fig:independent_high_speed_bl}, the present transformation gives the smallest buffer-layer envelopes among the three plotted transformations, while the outer portions retain stronger case dependence.  The corresponding values in table \ref{tab:buffer_spread} show that \(\Delta_{\mathrm{buf}}\) is smallest for the present transformation in both Cogo groups, with clearer separation in the \(M_\infty=2\)-4 group.

\begin{table}
  \begin{center}
  \small
  \setlength{\tabcolsep}{4.8pt}
  \begin{tabular}{L{0.40\textwidth}ccc}
  \hline
  Profile set & Trettel-Larsson & GFM & Present \\
  \hline
  Fig. \ref{fig:tbl_validation}(a-c), Zhang-Duan-Choudhari flat plates & 0.75 & 0.90 & 0.44 \\
  Fig. \ref{fig:tbl_validation}(d-f), Mach 2.28 station sequence & 1.25 & 0.99 & 1.06 \\
  Fig. \ref{fig:independent_high_speed_bl}(a), Cogo \(M_\infty=2\)-4 & 0.48 & 0.72 & 0.35 \\
  Fig. \ref{fig:independent_high_speed_bl}(b), Cogo \(M_\infty\simeq5.86\)-6 & 0.26 & 0.85 & 0.23 \\
  \hline
  \end{tabular}
  \caption{Buffer-layer envelope width \(\Delta_{\mathrm{buf}}\) for the boundary-layer profile sets in figures \ref{fig:tbl_validation} and \ref{fig:independent_high_speed_bl}.  Values are the mean difference between the upper and lower envelopes of the plotted profiles over \(5\le Y^+\le70\), after interpolation to a common \(Y^+\) grid.  Smaller values denote tighter profile collapse within each plotted set.}
  \label{tab:buffer_spread}
  \end{center}
\end{table}

\subsection{Shock-interaction departure and recovery}
\label{sec:sbli_recovery}

After the attached ZPG cases, the cooled-wall shock/boundary-layer-interaction (SBLI) DNS of \citet{HiraiKawai2023} introduces a non-equilibrium wall layer with streamwise-varying wall data.  The wall-surface data give \(\rho_w(x)\) and the skin-friction coefficient \(C_f(x)\).  The positive wall-shear magnitude used to define the wall friction velocity is
\begin{equation}
  \tau_w(x)=\max\left(|C_f(x)|/2,\tau_{\min}\right),
  \qquad
  u_{\tau,w}(x)=\left[\tau_w(x)/\rho_w(x)\right]^{1/2},
  \label{eq:sbli_tau_ref}
\end{equation}
Here the nondimensionalisation of \citet{HiraiKawai2023} is used, and \(\tau_{\min}=10^{-10}\).  Among the selected station profiles the smallest \(|C_f|/2\) is approximately \(7\times10^{-5}\), so the lower bound does not affect the plotted stations; it only specifies the limiting value of the reference friction velocity as the wall shear approaches zero.  The sign of \(C_f\) is retained when the wall-shear history and possible separation are interpreted, but the positive magnitude in (\ref{eq:sbli_tau_ref}) is used to form \(u_{\tau,w}\).  The boundary data for the reference-field equations are \(a_\rho=\log(\rho_w/\rho_0)\) and \(a_u=\log(u_{\tau,w}/u_{\tau,0})\) along the wall.  The shock produces an upstream pressure-gradient influence, a rapid change of wall scaling and a downstream recovery of wall shear, so the transformed profiles depart from attached-flow behaviour through the interaction and recover downstream.

\begin{figure}
  \centerline{\includegraphics[width=0.98\textwidth]{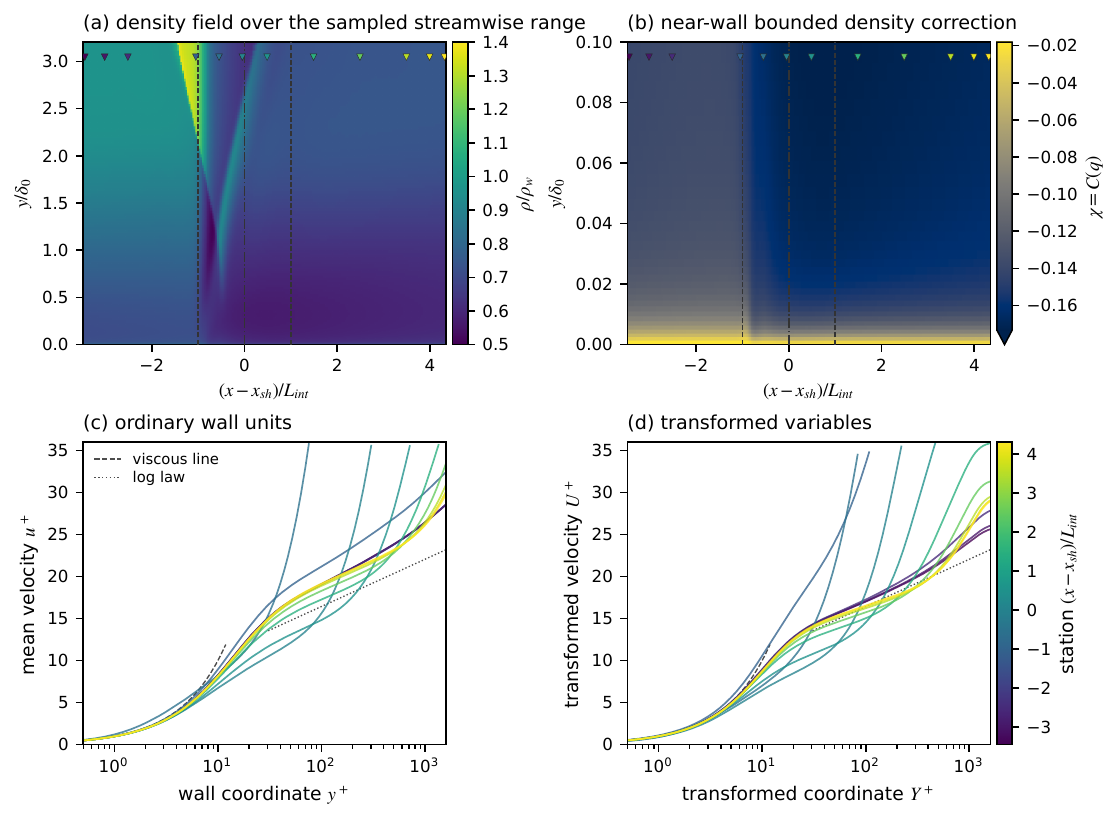}}
  \caption{Reference-field transformation through the cooled-wall shock/boundary-layer interaction of \citet{HiraiKawai2023}.  (a) Mean density normalised by the local wall value, $\rho/\rho_w(x)$, over $(x-x_{sh})/L_{int}\simeq -3.5$ to $4.3$, where \(x_{sh}\) is the nominal shock-impingement location and \(L_{int}\) is the interaction length used by \citet{HiraiKawai2023}; the vertical dashed bounds bracket the interaction and early recovery region, and the dash-dotted line marks shock impingement.  (b) Near-wall bounded density correction \(\chi=C(q)\) over the same streamwise range and \(0\le y/\delta_0\le0.10\), with the colour scale shown in the panel.  The wall values of density and friction velocity supply the Dirichlet data for \(a_\rho\) and \(a_u\).  Triangular markers in (a,b) denote the station profiles shown in (c,d).  (c) Mean velocity profiles in ordinary wall units $(y^+,u^+)$.  (d) Transformed profiles $(Y^+,U^+)$.  The colour of each profile matches its station marker and the station colour bar in (d), moving from upstream influence through the interaction and recovery region in increasing $(x-x_{sh})/L_{int}$.}
  \label{fig:sbli_recovery}
\end{figure}

Figure \ref{fig:sbli_recovery} shows the two-dimensional density field and near-wall correction over $(x-x_{sh})/L_{int}\simeq -3.5$ to $4.3$.  The density field identifies the cooled-wall interaction, with vertical dashed bounds bracketing the interaction and early recovery around shock impingement.  The bounded correction \(\chi=C(q)\) remains concentrated near the wall because it follows the local density contrast relative to the solved reference density.  The profiles in panels (c,d) are sampled from the solved \(Y^+\) and \(U^+\) fields.

In figure \ref{fig:sbli_recovery}(c,d), the selected stations cover the upstream pressure-gradient influence, the interaction region and the downstream recovery.  In ordinary wall units, the profiles first leave the far-upstream inner-layer scaling as the adverse-pressure-gradient influence reaches the wall.  The upstream mismatch in the transformed variables reflects adverse-pressure-gradient and wall-shear-history effects in addition to local density and viscosity stretching.  The interaction stations remain displaced, consistent with a separated or near-separated non-equilibrium layer.  Downstream of the interaction, the profiles for $(x-x_{sh})/L_{int}\gtrsim 1.5$ return toward a narrower transformed inner-layer band and toward the log law.  The station near $(x-x_{sh})/L_{int}\simeq -0.55$, where \(|C_f|\) is smallest among the selected profiles, lies in the interaction range and is the station most affected by the wall-shear normalisation in (\ref{eq:sbli_tau_ref}).

\section{Compressible channel flows}
\label{sec:two_wall_channels}

\subsection{Symmetric isothermal channels}
\label{sec:symmetric_channel}

Channels with equal isothermal walls provide a two-wall comparison in which the wall boundary data on both sides coincide.  For the symmetric isothermal channels of \citet{HuangColemanBradshaw1995} and \citet{GerolymosVallet2023,GerolymosVallet2024}, the reference-field equations reduce to
\[
  \rho_r=\rho_w,\qquad u_{\tau,r}=u_{\tau,w},\qquad
  S_\rho=\log(\rho/\rho_w),
\]
with the same logarithmic density term \(q=S_\rho/2\) and bounded correction \(\chi=C(q)\) used in the boundary-layer figures.  The two channel halves are statistically equivalent, so the plotted profiles are referenced to one wall.

\begin{figure}
  \centerline{\includegraphics[width=0.98\textwidth]{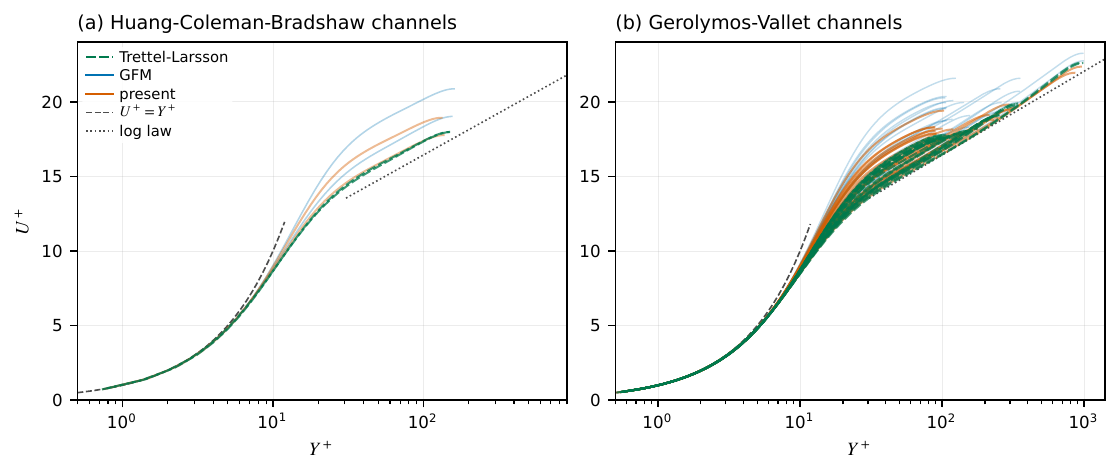}}
  \caption{Symmetric isothermal channel comparison.  (a) Huang-Coleman-Bradshaw channel profiles \citep{HuangColemanBradshaw1995}.  (b) Gerolymos-Vallet channel profiles \citep{GerolymosVallet2023,GerolymosVallet2024}.  Trettel-Larsson (green dashed), GFM (blue solid) and the present transformation (orange solid) are plotted as transformed velocity versus transformed coordinate.  The black dashed and dotted curves denote the viscous law and logarithmic law.  The comparison tests the density correction in symmetric isothermal channel flows.}
  \label{fig:symmetric_channel}
\end{figure}

Figure \ref{fig:symmetric_channel} compares the Huang-Coleman-Bradshaw and Gerolymos-Vallet profiles with Trettel-Larsson, GFM and the present transformation.  In both data families the present profiles remain in a compact inner-layer band and follow the logarithmic trend through the plotted overlap region.  Trettel-Larsson remains tighter in parts of the Gerolymos-Vallet set, consistent with the symmetric channel setting in which it was developed.  The figure shows how the present density correction behaves in symmetric isothermal channel data, while retaining Trettel-Larsson as the tighter comparator for this particular channel family.

\subsection{Mixed thermal-wall channels}
\label{sec:channel_gauge}

The Lusher-Coleman channels depart from this symmetric isothermal configuration by imposing an isothermal wall on one side and an adiabatic wall on the other \citep{LusherColeman2023}.  For these cases the reference fields are solved across the full channel, so that both wall boundary values enter before wall-referenced profiles are extracted.  Let \(y_l\) and \(y_u\) denote the lower and upper wall positions, and let \(h\) denote the channel half-height.  The logarithmic reference fields satisfy
\begin{equation}
  \begin{aligned}
  \frac{\dd^2a_\rho}{\dd y^2}&=0,&
  a_\rho(y_l)&=\log(\rho_{w,l}/\rho_0),&
  a_\rho(y_u)&=\log(\rho_{w,u}/\rho_0),\\
  \frac{\dd^2a_u}{\dd y^2}&=0,&
  a_u(y_l)&=\log(u_{\tau,l}/u_{\tau,0}),&
  a_u(y_u)&=\log(u_{\tau,u}/u_{\tau,0}),
  \end{aligned}
  \label{eq:srho_channel_wall}
\end{equation}
where \(\rho_{w,l}\), \(\rho_{w,u}\), \(\tau_{w,l}\) and \(\tau_{w,u}\) are the lower- and upper-wall density and wall-shear-stress magnitudes, with \(u_{\tau,l}=(\tau_{w,l}/\rho_{w,l})^{1/2}\) and \(u_{\tau,u}=(\tau_{w,u}/\rho_{w,u})^{1/2}\).  The physical reference fields, shear scale and inverse viscous length are then given by (\ref{eq:reference_fields}) and (\ref{eq:g_def}).  Since the two walls generally impose different values, the channel fields retain the two wall states before either wall profile is extracted.

The density contrast remains tied to the solved density reference,
\begin{equation}
  S_\rho(y)=\log[\rho(y)/\rho_r(y)] ,
  \label{eq:channel_wall_source}
\end{equation}
so \(\rho_r\), \(u_{\tau,r}\), \(\tau_r\), \(G\) and \(G_Y=G\exp\chi\) are channel fields defined by the two wall boundary values and the local density field.

The plotted lower- and upper-wall profiles are then extracted from the same corrected stretching and mean viscous shear, but each profile is referenced to its own wall.  The wall-tangential viscous shear is taken positive in the wall-normal direction away from the corresponding wall and is scaled by \(\tau_r=\rho_r u_{\tau,r}^2\).  Thus the channel fields \(\rho_r\), \(u_{\tau,r}\), \(\tau_r\), \(\chi\), \(G\) and \(G_Y\) are common to the two-wall problem, while the plotted pairs \((Y_l^+,U_l^+)\) and \((Y_u^+,U_u^+)\) retain their own wall origins and outward wall-normal directions; the wall friction velocities enter through the boundary data for \(u_{\tau,r}\).

\begin{figure}
  \centerline{\includegraphics[width=0.98\textwidth]{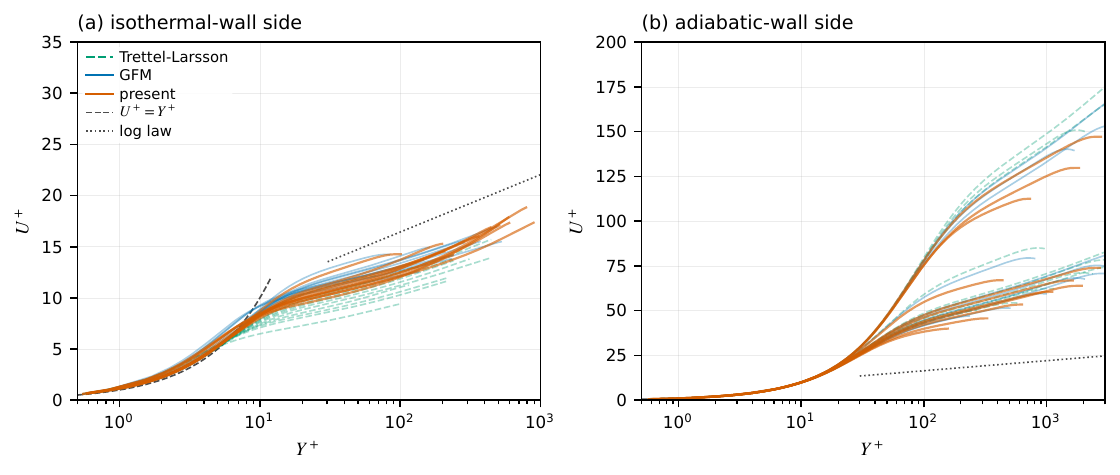}}
  \caption{Mixed thermal-wall channel comparison for the Lusher-Coleman cases \citep{LusherColeman2023}.  (a) Profiles referenced to the isothermal wall.  (b) Profiles referenced to the adiabatic wall, shown over the full plotted range to expose the outer displacement.  Trettel-Larsson (green dashed), GFM (blue solid) and the present transformation (orange solid) are plotted as transformed velocity versus transformed coordinate.  For the present curves, the full-channel reference fields are obtained from (\ref{eq:srho_channel_wall}) and the density contrast is \(S_\rho=\log(\rho/\rho_r)\), with \(q=S_\rho/2\) and \(\chi=C(q)\).  The black dashed and dotted curves denote the viscous law and logarithmic law.}
  \label{fig:mixed_channel_profiles}
\end{figure}

Figure \ref{fig:mixed_channel_profiles} compares the resulting present profiles with Trettel-Larsson and GFM for the same mixed-channel cases.  On the isothermal-wall side the three transformations remain close through the inner and buffer layers.  On the adiabatic-wall side they all retain a large outer displacement, showing the side-dependent residual left by unequal wall thermal states.

\section{Discussion and range of use}
\label{sec:discussion}

The uniform-wall boundary layers isolate the density correction from the reference-field part of the formulation.  In these cases the reference fields reduce to the wall values, and ordinary semi-local scaling already supplies the leading viscous-length change.  The remaining action is the bounded half-density correction in (\ref{eq:kinematic_viscosity_limit}).  Figures \ref{fig:collapse_overview} and \ref{fig:tbl_validation}, together with table \ref{tab:buffer_spread}, show that this correction is most visible in the cooled Zhang-Duan-Choudhari cases, where the density increase is strong and concentrated in the inner layer.  In the developing Mach 2.28 set, GFM and the present transformation have comparable envelopes because the density correction is a modest addition to the ordinary semi-local stretching.  In the Cogo set, the present transformation still gives the smallest envelopes in table \ref{tab:buffer_spread}, but the separation is limited in the high-Mach group where Trettel-Larsson is already compact.

Figures \ref{fig:field_patch} and \ref{fig:sbli_recovery} show how the field formulation enters the sampled profiles.  In the Mach 2.28 boundary-layer patch, \(Y^+\) and \(U^+\) are defined over the wall-layer domain and station profiles are sections through those fields.  In the shock-interaction case, the wall density and friction velocity vary along the surface and enter the reference fields as boundary data.  The transformed profiles record both the departure from the attached-flow band under adverse pressure gradient and the downstream recovery as the wall shear increases again.  The station with the smallest selected \(|C_f|\) remains a limit case for any friction-scaled transformation, as indicated in figure \ref{fig:sbli_recovery}.

The channel cases distinguish symmetric isothermal walls from mixed thermal wall conditions.  Figure \ref{fig:symmetric_channel} gives the symmetric isothermal comparison: the present transformation gives a compact family, but Trettel-Larsson is still tighter for part of the Gerolymos-Vallet set.  Figure \ref{fig:mixed_channel_profiles} then introduces unequal thermal wall states.  The isothermal-wall profiles remain close among the three transformations, whereas the adiabatic-wall profiles retain a larger outer displacement.  This residual is consistent with two wall layers that do not share a single thermal state or friction scale; a single local coordinate stretching cannot make both sides behave as the same canonical inner layer.

The DNS comparisons point to the flow conditions in which the correction is most visible.  The favourable case is an attached or recovering wall layer with strong near-wall density and viscosity variation, especially under wall cooling.  Weak property variation, late hot-wall development and separated or nearly separated regions leave smaller or less reliable gains.  Mixed thermal-wall channels mark a boundary of the profile collapse: the full-channel reference fields retain both wall boundary conditions, but the extracted wall profiles still reflect the unequal thermal states.  Strongly three-dimensional wall temperature, blowing, ablation, reacting walls and more complicated multi-wall configurations remain outside the evidence considered here.

\section{Conclusions}
\label{sec:conclusions}

We formulate a semi-local transformation in which the transformed coordinate and velocity are defined by wall-anchored field equations before profiles are sampled.  The wall density and friction velocity enter through boundary conditions for auxiliary field equations, rather than through an algebraic assignment of wall values to interior points.  The resulting fields \(\rho_r\) and \(u_{\tau,r}\) set the inverse viscous length \(G=\sqrt{\rho\,\rho_r}u_{\tau,r}/\mu\), while the local density contrast \(S_\rho=\log(\rho/\rho_r)\) gives \(q=S_\rho/2\), the bounded correction \(\chi=C(q)\), and the corrected stretching \(G_Y=G\exp\chi\).  The transformed coordinate and velocity are then obtained from this corrected stretching and the mean viscous shear.  In the constant-property limit the construction reduces to the usual wall coordinate and velocity in wall units.

The clearest effect in the DNS comparisons occurs in cooled high-speed boundary layers, where the density contrast is large and concentrated in the inner layer.  For the Zhang-Duan-Choudhari profiles shown here, the transformation reduces the buffer-layer spread relative to ordinary wall scaling and to the Trettel-Larsson and GFM transformations.  The developing Mach 2.28 and Cogo boundary-layer cases apply the field-equation formulation outside the initial cooled flat-plate examples: the Mach 2.28 set gives a small distinction between GFM and the present transformation, while the Cogo set retains the smallest envelope for the present transformation with limited separation in the high-Mach group.  In symmetric isothermal channels, the same density correction gives a compact profile family, although Trettel-Larsson remains tighter for this particular channel set.  The mixed thermal-wall channels give a different outcome.  The reference fields can be solved across the full channel with both wall boundary values, but the extracted isothermal- and adiabatic-wall profiles retain a marked side dependence, especially on the adiabatic-wall side.

The shock/boundary-layer-interaction case illustrates the field-equation construction when wall density and friction velocity vary along the wall.  The transformed profiles move away from the attached-flow inner-layer band under the adverse-pressure-gradient and interaction influence, and move back toward it as the wall shear recovers downstream.  Stations with very small \(|C_f|\) remain sensitive to wall-shear normalisation and should not be interpreted as attached-flow collapse through the separated part of the interaction.  Strongly three-dimensional wall temperature, blowing, ablation, reacting walls and multiple interacting walls remain open cases beyond the present evidence.

\backsection[Acknowledgements]{Zifei Yin thanks Prof. Guowei He and Prof. Paul Durbin for their support and encouragement in exploring statistical theory and modelling of turbulent flows throughout the years.}

\backsection[Funding]{This work was supported by the National Natural Science Foundation of China (No. 12472226 and 92052203).}

\backsection[Declaration of interests]{The authors report no conflict of interest.}

\backsection[Data availability statement]{The DNS datasets used in this manuscript are available from the public archives cited in \S\ref{sec:data}.  The flat-plate cases labelled by $M_\infty$ and wall-temperature ratio are the Zhang-Duan-Choudhari DNS statistics, including cases where the same statistics appear in more than one public archive.}

\backsection[Use of AI tools]{AI tools were used for limited assistance with language editing, bibliographic review and locating cited public DNS datasets. The authors remain responsible for all scientific content, data analysis, and conclusions.}

\bibliographystyle{jfm}
\bibliography{references}

\end{document}